\documentclass[twoside]{article}

\usepackage[accepted]{arxiv}
\usepackage[round]{natbib}
\usepackage{answers}
\usepackage{setspace}
\usepackage{graphicx}
\usepackage{graphics}
\usepackage{subfigure}
\usepackage{enumitem}
\usepackage{multicol, multirow}
\usepackage{mathrsfs}
\usepackage{bbm}
\usepackage{amsmath,amsthm,amssymb}
\usepackage{rotating}
\usepackage{float}
\usepackage[hidelinks]{hyperref}
\usepackage{url}
\usepackage{mathtools}
\usepackage{caption}
\usepackage[ruled,vlined,noresetcount]{algorithm2e}
\usepackage[toc,page]{appendix}
\usepackage{color}
\usepackage{booktabs}       
\usepackage[export]{adjustbox}

\usepackage[dvipsnames]{xcolor}
\usepackage{tikz}
\usetikzlibrary{bayesnet}
\usetikzlibrary{math}
\usetikzlibrary{arrows.meta}
\usepackage{tikzscale}

\DeclarePairedDelimiterX{\infdivx}[2]{\Big[}{\Big]}{%
  #1\;\delimsize\|\;#2%
}

\newcommand{\vecs}{\mathbf{s}}
\newcommand{\vecx}{\mathbf{x}} 

\newcommand{\vecxi}{\pmb{\xi}}

\newcommand{\vecz}{\mathbf{z}}

\newcommand{\vectheta}{\pmb{\theta}}

\newcommand{\V}{\mathbf{V}}

\newcommand{\Lambdab}{\pmb{\Lambda}}
\newcommand{\Prob}{\mathbb{P}}
\newcommand{\prob}{p}

\newcommand{\Beps}{\mathrm{B}_{\epsilon}}
\newcommand{\prior}{\pi}
\newcommand{\post}{\pi} 
\newcommand{\proposal}{p}
\newcommand{\cheappost}{\tilde{\post}}
\newcommand{\auxpost}{p}
 
\newcommand{\obs}{\vecx^o}

\begin{document}

\twocolumn[

\aistatstitle{Adaptive Gaussian Copula ABC}

\aistatsauthor{ Yanzhi Chen \And Michael U. Gutmann}
\aistatsaddress{The University of Edinburgh \And The University of Edinburgh } ]

\begin{abstract}
  \mbox{Approximate Bayesian computation (ABC)} is a set of techniques for Bayesian inference when the likelihood is intractable but sampling from the model is possible. This work presents a simple yet effective ABC algorithm based on the combination of two classical ABC approaches --- regression ABC and sequential ABC. The key idea is that rather than learning the posterior directly, we first target another auxiliary distribution that can be learned accurately by existing methods, through which we then subsequently learn the desired posterior with the help of a Gaussian copula. During this process, the complexity of the model changes adaptively according to the data at hand. Experiments on a synthetic dataset as well as three real-world inference tasks demonstrates that the proposed method is fast, accurate, and easy to use. 
\end{abstract}

\section{Introduction}
Many parametric statistical models are specified in terms of a parametrised data generating process. We can sample or simulate data from this kind of models but their likelihood function is typically too costly to evaluate. The models are called implicit \citep{Diggle1984} or simulator-based models \citep{bo-abc} and are widely used in scientific domains including ecology \citep{example1},  epidemiology \citep{Corander2017}, psychology \citep{example4} and cosmology \citep{example3}. For example, the demographic evolution of two species can be simulated by a set of stochastic differential equations controlled by birth/predation rates but computation of the likelihood is intractable. Often, the true interest is rarely in simulating data, but in the inverse problem of identifying model parameters that could have plausibly produced the observed data. This usually not only includes point estimates of the model parameters but also a quantification of their uncertainty. 

The above task can, in principle, be solved by the well-known framework of Bayesian inference. However, due to the absence of likelihood functions for implicit models, exact inference becomes difficult or even impossible. The technique of approximate Bayesian computation (ABC) enables inference in such circumstances \citep[for recent reviews, see e.g.][]{Lintusaari2017, Sisson2018}. The most basic ABC algorithm works by repeatedly simulating data with different parameters and only accepting those parameters whose simulated data 'resemble' the observed ones. This generally requires hitting a small $\epsilon$-ball for good accuracy, which is unfortunately impractical in high-dimensional spaces due to the well-known curse of dimensionality \citep[see e.g.][]{Beaumont2010}. 

Several efforts have been made to alleviate the above accuracy-efficiency dilemma in ABC \citep[e.g.][]{mcmc-abc,linRegABC, smc-abc, nonlinRegABC, bo-abc, mdn-abc1}. The first line of work, regression ABC \citep{linRegABC, nonlinRegABC}, accelerates inference by first allowing a much larger $\epsilon$-ball, and then capturing the relationship between the simulated data and the parameters inside this $\epsilon$-ball through a regression function, with which accepted samples are post-adjusted so as to compensate for the accuracy loss due to the larger $\epsilon$ used. Another line of work improves the efficiency of ABC by learning the posterior iteratively \citep{smc-abc2, smc-abc, mdn-abc1, mdn-abc2}. In these methods, preliminary models of the posterior are used to guide further simulations, leading to a significant drop of overall simulation costs. Both lines of methods are shown to be effective in a wide range of inference tasks. 

In this paper, we show how these two lines of works can refine each other, yielding a powerful yet simple algorithm. We bridge the two lines of work using Gaussian copula modelling \citep{gc-abc,gc-abc2}. The resulting proposed approach, which we call adaptive Gaussian copula ABC (AGC-ABC), has similar power as recent flexible machine-learning methods \citep{mdn-abc1, mdn-abc2} but requires less simulated data. This naturally leads to performance gains for smaller computational budgets, as we confirm empirically.

The rest of the paper is organized as follows. Section 2 gives an overview of regression ABC and sequential ABC. Section 3 details the proposed AGC-ABC algorithm. Section 4 compares the proposed method to existing works and Section 5 contains the experimental results.  Section 6 concludes the paper. 
\section{Background}
Given observed data $\obs$, the task of approximate Bayesian inference for implicit models is to estimate the posterior density of the  parameters $\vectheta$ without evaluating the likelihood directly. To achieve this goal, methods known as approximate Bayesian computation (ABC) typically reduce the observed data to summary statistics $\vecs^o = \vecs(\obs)$ and approximate the likelihood by $\Prob(\|\vecs - \vecs^o\| < \epsilon | \ \vectheta)$ where $\epsilon$ is chosen sufficiently small. The approximate posterior is then
\begin{align}
    \post(\vectheta |\vecs^o) &\propto \pi(\vectheta) \cdot \Prob(\|\vecs - \vecs^o\| < \epsilon |\vectheta).
    \label{formula:abc-post}
\end{align}
ABC algorithms typically only yield samples from the above approximate posterior. The most basic algorithm, ABC rejection sampling, first proposes $\vectheta$ from the prior $\pi(\vectheta)$ and then only accepts those $\vectheta$ whose simulated summary statistics fall within an $\epsilon$-ball $\Beps(\vecs^o) = \{\vecs: \| \vecs - \vecs^o \| \le \epsilon\}$ around $\vecs^o$. While simple and robust, the algorithm suffers from a poor trade-off between accuracy and computational efficiency \citep[see e.g.][]{Lintusaari2017}. Two lines of work aim at improving the trade-off from different perspectives: regression ABC and sequential ABC. 

\subsection{Regression ABC}
This line of work first allows a much larger $\epsilon$-ball in rejection ABC, and then uses the accepted samples to estimate the relationship between $\vectheta$ and $\vecs$ through regression. The learned regression model is then used to adjust the samples so that, if the regression model is correct, the adjusted samples correspond to samples from $\post(\vectheta | \vecs^o)$ in \eqref{formula:abc-post} with $\epsilon = 0$.

Conditional on the summary statistics $\vecs$, an additive noise model is typically used to describe the relationship between $\vectheta$ and $\vecs$ as
\begin{equation}
\vectheta|\vecs = \quad  \underbrace{\mathbb{E}[\vectheta|\vecs] }_{\text{mean}} \quad  +  \underbrace{\quad \vecxi|\vecs \quad}_{\text{residual}}.
\label{formula:additive}
\end{equation}
Here, we used the notation $\vectheta|\vecs$ and $\vecxi|\vecs$ to highlight that the above relationship is conditional on $\vecs$. In particular, the distribution of the residual may depend on the value of $\vecs$ so that $p(\vecxi | \vecs) \neq p(\vecxi)$. Under the above model, to sample from the posterior $\post(\vectheta | \vecs^o)$, we need to evaluate $\mathbb{E}[\vectheta|\vecs^o]$ and sample from the conditional distribution of $\vecxi$ at $\vecs^o$.

To learn $\mathbb{E}[\vectheta|\vecs]$, regression ABC fits a regression function $g(\cdot)$ on the samples $(\vectheta^{(i)}, \vecs^{(i)})$, $i=1, \ldots, I,$ collected in the preliminary run of rejection ABC. The fitting is typically done by the method of least squares, i.e.\ by minimising
\begin{equation}
    J(g) = \frac{1}{I} \sum_{i=1}^I  \| \vectheta^{(i)} - g(\vecs^{(i)}) \|^2_2 
    \label{formula:square-loss}
\end{equation}
since it is simple and, for sufficiently many samples (training data), the optimal $g(\vecs)$ is guaranteed to approach $\mathbb{E}[\vectheta|\vecs]$. A critical aspect of this approach is the choice of the function family among which one searches for the optimal $g$. For small amounts of training data, linear regression \citep{linRegABC} is a safe choice while more flexible neural networks \citep{nonlinRegABC} may be used when more training data are available.

On obtaining $g(\cdot)$ we get an approximation for the posterior mean at $\vecs^o$ via $g(\vecs^o)$. What remains unknown is the distribution $p(\vecxi|\vecs^o)$. To estimate this, regression ABC makes the \emph{homogeneity} assumption that the distribution of the residual remains roughly unchanged for all $\vecs$ in the $\epsilon$-ball $\Beps(\vecs^o)$ around $\vecs^o$:
\begin{equation}
    p(\vecxi|\vecs) \approx p(\vecxi) \qquad \forall \vecs \in \Beps(\vecs^o)
    \label{formula:homo}
\end{equation}
With this assumption, the empirical residuals $\vectheta^{(i)} - g(\vecs^{(i)})$ are used as samples from $p(\vecxi | \vecs^o)$, and the adjustment
\begin{equation}
\vectheta'^{(i)} = \underbrace{g(\vecs^o)}_{\text{est.mean}} + \quad \underbrace{\vectheta^{(i)} - g(\vecs^{(i)})}_{\text{est.residual}}
\label{formula:adjustment}
\end{equation}
yields samples from $\post(\vectheta | \vecs^o)$ in \eqref{formula:abc-post} with $\epsilon = 0$. However, this result hinges on (a) the mean function being learned well and (b) the homogeneity assumption holding so that the estimated residuals do correspond to samples from $p(\vecxi|\vecs^o)$.
For a theoretical analysis of regression ABC, see e.g\ \citep{ABCnonPara}.

\subsection{Sequential ABC}
This second line of work aims at accelerating the inference by learning the posterior in an iterative way \citep{smc-abc,smc-abc2,mdn-abc1,mdn-abc2}. Using the prior $\prior(\vectheta)$ as proposal distribution in the generation of the parameter-data tuples $(\vectheta^{(i)}, \vecs^{(i)})$, the methods first learn a rough but computationally cheaper approximation $\cheappost(\vectheta | \vecs^o)$ to the posterior, e.g.\ by using a larger $\epsilon$ value \citep{smc-abc, smc-abc2}, or a small fraction of the overall simulation budget \citep{mdn-abc1, mdn-abc2}. The learned posterior $\cheappost(\vectheta | \vecs^o)$ is then used as a new proposal distribution (new 'prior') $\proposal(\vectheta)$ in the next round. From newly generated tuples $(\vectheta^{(i)}, \vecs^{(i)})$, with $\vectheta^{(i)} \sim \proposal(\vectheta)$, the methods then construct an auxiliary posterior 
$\auxpost(\vectheta|\vecs^o)$. In order to correct for using another proposal distribution than the prior, the auxiliary posterior $\auxpost(\vectheta|\vecs^o)$ and samples from it need to be reweighted by
\begin{equation}
    w(\vectheta) \propto \frac{\pi(\vectheta)}{\proposal(\vectheta)}
    \label{formula:recover}
\end{equation}
see e.g.\ \citep{Lintusaari2017}. A reason why this method accelerates inference is that the new, more informed, proposal $\proposal(\vectheta)$ generates more often summary statistics $\vecs$ that fall inside the $\epsilon$-ball $\Beps(\vecs^o)$. For a given $\epsilon$, this then results in a higher acceptance rate and a reduction of the simulation cost. 

The concrete algorithms of sequential ABC may differ across methods. Earlier sequential Monte Carlo ABC methods \citep[SMC-ABC:][]{smc-abc, smc-abc2} adopt a non-parametric posterior modeling, whereas the recent sequential neural posterior ABC ones \citep[SNP-ABC:][]{mdn-abc1, mdn-abc2} take a parametric approach using mixture density networks. However, the underlying basic principles remain roughly the same. 

\begin{figure*}[t!]
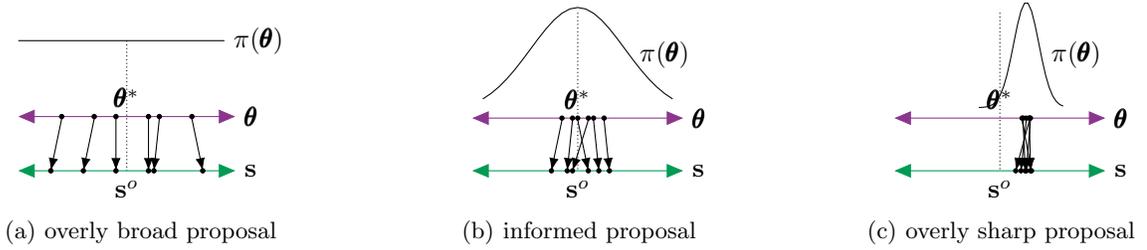

    \vspace{0.5cm}
    \centering
    \hspace{-0.020\linewidth}
    \subfigure[overly broad proposal]{
        \includegraphics[width=0.235\linewidth]{figures/a.tikz}
    }
    %\hspace{0.035\linewidth}
    \hspace{0.09\linewidth}
     \subfigure[informed proposal]{        
        \includegraphics[width=0.21\linewidth]{figures/b.tikz}
    }
    %\hspace{0.035\linewidth}
    \hspace{0.09\linewidth}
    \subfigure[overly sharp proposal]{        
        \includegraphics[width=0.21\linewidth]{figures/c.tikz}
    }
    \caption{The effect of different proposal distributions in regression ABC.}
    \label{figure:agc-theory}
\end{figure*}

\section{Methodology}
We here explain the proposed approach that makes use of the basic principles of both regression ABC and sequential ABC.

The performance of regression ABC is likely to suffer if the homogeneity assumption in \eqref{formula:homo} does not hold. If ABC is run with a broad proposal distribution (e.g.\ the prior), only very few of the simulated $\vecs$ are close to $\vecs^o$, so that deviation from the homogeneity is generally likely (Figure \ref{figure:agc-theory}.a). However, if we use a more informed proposal distribution, more samples of $\vecs$ around $\vecs^o$ can be collected (Figure \ref{figure:agc-theory}.b). In other words, the shape of the proposal distribution strongly affects whether the homogeneity assumption in regression ABC reasonably holds. This motivates us to take a two-stage approach where, in a first coarse-grained phase, we learn a rough approximation to the posterior and then, in a subsequent fine-grained-phase, use it as proposal distribution in order to generate a sufficient amount of $(\vectheta, \vecs)$ pairs for which the homogeneity assumption in regression ABC is better satisfied. 

\textbf{Regression function}. We will be using regression-ABC in both the coarse- and fine-grained phase, for which we need to choose a family of regression functions $g$. Simple regression models with linear functions $g$ are easy to learn but are not able to capture possible nonlinear relationships between $\vecs$ and $\vectheta$. On the other hand, neural networks are more flexible; however they also require more training data. Here, we thus perform model selection by comparing the validation error between linear regression and a neural network trained with early stopping. The split into training/validation set is done on a 80/20 basis. 
Before regression, each $\vectheta$ is preprocessed such that they have the same scale along each dimension.

\textbf{Coarse-grained phase}. In this stage we learn a rough but robust approximation $\cheappost(\vectheta | \vecs^o)$ to the posterior using only a small fraction $\lambda$ of the overall simulation budget $N$. We do this by performing ordinary regression ABC based on parameters generated from the prior $\prior(\vectheta)$. We then pick a subset of the adjusted samples $\{\vectheta'^{(1)},...,  \vectheta'^{(m)}\}$ whose corresponding summary statistics $\vecs$ are among the top $m$ closest to $\vecs^o$ and model the adjusted samples with a Gaussian distribution so that
\begin{equation}
    \cheappost(\vectheta | \vecs^o) = \mathcal{N} \Big(\vectheta; g(\vecs^o), \V \Big).
\label{formula:proposal-prior}
\end{equation}
Here, $g(\cdot)$ is the learned regression function and $\V$ is the inflated sample covariance matrix computed as $\V = \alpha \cdot \frac{1}{m} \sum_{i}^{m}  [\vectheta'^{(i)} - g(\vecs^o) ][\vectheta'^{(i)} - g(\vecs^o) ]^\top$, $\alpha=1.5$. By using a Gaussian, we work with the maximum entropy distribution given mean and covariance of the adjusted samples. The motivation for inflating the covariance matrix is that a slightly broader proposal distribution is more robust to e.g.\ estimation errors in the learned regression function (Figure \ref{figure:agc-theory}.c).

For the value of $m$, we choose $m$ so that the top 20\% summary statistics closest to $\vecs^o$ are retained, which is a conservative but robust strategy given the small simulation budget in this phase. The whole coarse-grained learning procedure is summarized in Algorithm 1, where, for simplicity, we incorporated the reduction to summary statistics directly into the data generating process (simulator). 

\textbf{Fine-grained phase}. After learning $\cheappost(\vectheta | \vecs^o)$ we use it as proposal distribution $\proposal(\vectheta)$ in another round of regression ABC with the remaining  $1-\lambda$ simulation budget. We adjust the top $n$ samples whose summary statistics $\vecs$ are closest to $\vecs^o$ and then model their distribution as a Gaussian copula. 

Gaussian copula are a generalisation of the Gaussian distribution. They can be seen to model (the adjusted) $\vectheta' = (\theta'_1, \ldots, \theta'_K)^\top$ in terms of latent variables $\vecz$ that are element-wise non-linearly transformed to give $\theta'_k = F_k^{-1}(\Phi(z_k))$ where $F_k$ is the marginal cumulative distribution function (CDF) of $\theta'_k$ and $\vecz \sim \mathcal{N}(\vecz; \mathbf{0}, \Lambdab)$. Under this construction, the auxiliary posterior $\auxpost(\vectheta | \vecs^o)$ of the adjusted samples equals
\begin{equation}
    \auxpost(\vectheta | \vecs^o) = \underbrace{\frac{\mathcal{N}(\vecz; \mathbf{0}, \Lambdab)}{\prod \mathcal{N}(z_k; 0, 1)}}_{c_{\text{GC}}(u_1,...,u_k; \Lambda)} \cdot \Bigg[ \prod \limits^K_k f_k(\theta_k) \Bigg],
    \label{formula:gc-def}
\end{equation}
where $\Lambdab$ is the correlation matrix whose diagonal elements are all ones and $c_{\text{GC}}(\cdot)$ is called the Gaussian copula density. 

The learning of Gaussian copula can be done in a semi-parametric way. Denote the adjusted samples by $\{\vectheta'^{(1)},...,  \vectheta'^{(n)}\}$:
\begin{itemize}[leftmargin=*]
    \item \emph{Marginal distribution}. Each $f_k(\theta_k)$ is learned by kernel density estimation (KDE), which is well-known to be typically accurate in 1D \citep{kde}.
    \item \emph{Correlation matrix}. The matrix $\Lambdab$ is learned by converting each sample $\vectheta^{'(j)}$ to its corresponding latent representation $\vecz^{(j)}$ as $z^{(j)}_k = \Phi^{-1}(r_k(\theta^{'(j)}_k))$ where  $r_k(x) = \frac{1}{n} \sum^n_{j=1} \mathbf{1}[\theta^{'(j)}_k < x]$, followed by estimating $\Lambdab$ as $\Lambdab = \frac{1}{n}\sum_j^n \vecz^{(j)}\vecz^{(j)\top}$.
\end{itemize}

As for sequential ABC in \eqref{formula:recover}, we need to convert the auxiliary posterior $\auxpost(\vectheta|\vecs^o)$ to an estimate of the actual posterior $\post(\vectheta | \vecs^o)$ by reweighing it by the ratio of the prior to proposal distribution. This gives
\begin{align}
    \post(\vectheta | \vecs^o) \propto \frac{\prior(\vectheta)}{\cheappost(\vectheta | \vecs^o)} \auxpost(\vectheta | \vecs^o),
    \label{formula:agc-post}
\end{align}
where the proposal distribution $\cheappost(\vectheta | \vecs^o)$ was defined in \eqref{formula:proposal-prior} and $\auxpost(\vectheta | \vecs^o)$ was defined in \eqref{formula:gc-def}. Note that while the auxiliary posterior $\auxpost(\vectheta|\vecs^o)$ is a Gaussian copula, the re-weighted $\post(\vectheta | \vecs^o)$ is generally not (because of e.g.\ the presence of the prior). 

Unlike in the coarse-grained phase, in this phase, we select $n$ such that the top 2000 summary statistics closest to $\vecs^o$ are retained. We adopt this more aggressive strategy because here we have a larger simulation budget and better satisfy the homogeneity condition. We are therefore safe to keep a larger number of samples so as to reduce the Monte Carlo error. In addition, by retaining a fixed number of samples we effectively shrink the value of $\epsilon$ when the overall simulation budget $N$ is increased. The whole fine-grained phase is summarized in Algorithm 2.

\section{Comparisons to other methods}
Here we compare AGC-ABC with three methods in the field: neural network regression ABC \citep[NN-ABC:][]{nonlinRegABC}, Gaussian copula ABC \citep[GC-ABC:][]{gc-abc}, and sequential neural posterior ABC \citep[SNP-ABC:][]{mdn-abc1, mdn-abc2}. 

\textbf{AGC-ABC vs.\ NN-ABC}. Both AGC-ABC and NN-ABC address the homogeneity issue (\ref{formula:homo}) in regression ABC. NN-ABC aims to model possible heterogeneity of the residuals by taking into account the possibly changing scale of the residuals. In contrast to this, AGC-ABC uses a two-stage approach to stay in the homogeneous regime, where not only the scale but also the dependency structure of the residuals is modelled.

\textbf{AGC-ABC vs.\ GC-ABC}. Both AGC-ABC and GC-ABC make use of Gaussian copula in posterior modeling. GC-ABC can be viewed as a special case of AGC-ABC without the adapting coarse-grained stage. The posterior in GC-ABC is a Gaussian copula whereas ours is generally not.

% Algorithm 1
\begin{algorithm}
    \SetAlgoNoLine
    \textbf{Input:} prior $\pi(\vectheta)$, observed data $\vecs^o$, simulation budget $\lambda N$, simulator $M(\vectheta)$ \\
    \textbf{Output:} coarse approximation $\cheappost(\vectheta | \vecs^o)$ \\
    \textbf{Hyperparams:} $m = \lceil 0.2\lambda N \rceil$. \\

    \quad \\

    \For{$i$ = $1$ $\text{to}$ $\lceil \lambda N \rceil$}{
        sample $\vectheta^{(i)} \sim \pi(\vectheta)$ \;
        simulate $\vecs^{(i)} \sim  M(\vectheta)$  \;
    }
    fit $g$ = $\arg \min_{\tilde{g}} J(\tilde{g})$ by \eqref{formula:square-loss} with $\{(\vectheta^{(i)}, \vecs^{(i)})\}^{m}_{i=1}$  \;
    $\vectheta^{'(i)} \leftarrow g(\vecs^o) + \vectheta^{(i)} - g(\vecs^{(i)})$, for $\forall i$\;
    sort $\vectheta^{'(i)}$ by $ \| \vecs^{(i)} - \vecs^o \|$ in ascendant order \;
    \quad \\
    $\pmb{\mu} \leftarrow g(\vecs^o)$ \;
    $\mathbf{V} \leftarrow \frac{3}{2m} \sum_{i=1}^{m}{(\vectheta'^{(i)}-\pmb{\mu})} {(\vectheta'^{(i)}-\pmb{\mu})}^{\top}$ \;
    construct $\cheappost(\vectheta | \vecs^o)$ with $(\pmb{\mu}, \mathbf{V})$ as in (\ref{formula:proposal-prior}) \;
    return $\cheappost(\vectheta | \vecs^o)$.% in \eqref{formula:proposal-prior}. 
    \caption{\texttt{Coarse-grained learning}}
\end{algorithm}

\textbf{AGC-ABC vs.\ SNP-ABC}. Both AGC-ABC and SNP-ABC learn the posterior in a coarse-to-fine fashion. The difference lies in the model complexity. SNP-ABC models the relationship between $\vectheta$ and $\vecs$ with a mixture of Gaussians whose parameters are computed by a neural network. While such constructions can approximate the posterior more and more accurately as the number of mixture components increases, the number of neural network parameters is generally much larger than in AGC-ABC where we only need to estimate the posterior mean and a correlation matrix. Additionally, the marginal distributions in our copula-based method can generally be modelled more accurately than in SNP-ABC, since we learn them nonparametrically for each dimension.

\section{Experiments}
\subsection{Setup}
We compare the proposed adaptive Gaussian Copula ABC (AGC-ABC) with five ABC methods in the field: rejection ABC \citep{RejABC}, regression ABC \citep{linRegABC}, GC-ABC \citep{gc-abc}, NN-ABC \citep{nonlinRegABC} and SNP-ABC \citep{mdn-abc1}. Comparisons are done by computing the JSD between the true and the approximate posterior, and when the true posterior is not available, we approximate it by running a ABC rejection sampling algorithm with (a) well chosen summary statistics $\vecs$ and (b) an extremely small $\epsilon$ value e.g the $0.00001$ quantile of the population of $\|\vecs - \vecs^o \|$, and then estimate it by KDE. Similarly, when the posterior approximated in each method is not available in analytical form (e.g for REJ-ABC, REG-ABC and NN-ABC), we estimate it by KDE. For two distributions $P$ and $Q$, the JSD is given by
\begin{align*}
  \mathrm{JSD}(P, Q) &= \frac{1}{2}\text{KL}\Big[P \| \frac{P+Q}{2} \Big] + \frac{1}{2}\text{KL}\Big[Q \| \frac{P+Q}{2} \Big]. 
\end{align*}

% Algorithm 2
\begin{algorithm}
    \SetAlgoNoLine
    \textbf{Input:} prior $\pi(\vectheta)$, proposal $\cheappost(\vectheta | \vecs^o)$, observed data $\vecs^o$, simulation budget $(1-\lambda) N$, simulator $M(\vectheta)$  \\
    \textbf{Output:} estimated posterior $\post(\vectheta|\vecs^o)$ \\
    \textbf{Hyperparams:} $n = 2000$. \\
    
    \quad \\

    \For{$i$ = $1$ $\text{to}$ $\lfloor (1-\lambda)N \rfloor$}{
        sample $\vectheta^{(i)} \sim \cheappost(\vectheta | \vecs^o)$ \;
        simulate $\vecs^{(i)} \sim  M(\vectheta)$  \;
    }
    fit $g$ = $\arg \min_{\tilde{g}} J(\tilde{g})$ by \eqref{formula:square-loss} with $\{(\vectheta^{(i)}, \vecs^{(i)})\}^{n}_{i=1}$  \;
    $\vectheta^{'(i)} \leftarrow g(\vecs^o) + \vectheta^{(i)} - g(\vecs^{(i)})$, for $\forall i$\;
    sort $\vectheta^{'(i)}$ by $ \| \vecs^{(i)} - \vecs^o \|$ in ascendant order \;
    \quad \\
    $f_k(\theta) \leftarrow $ \texttt{KDE}($\{ \theta_k^{'(1)},..., \theta_k^{'(n)} \}$), $\forall k$  \;
    $\mathbf{\Lambda} \leftarrow \frac{1}{n} \sum_{i=1}^{n}\vecz^{(i)}\vecz^{(i)\top}$ with $z_k^{(i)} = \Phi^{-1}(r_k(\theta_k^{'(i)}))$ \;
    construct $\auxpost(\vectheta|\vecs^o)$ with $(\{f_k(\theta_k)\}^K_{k=1}, \mathbf{\Lambda})$ as in (\ref{formula:gc-def}) \;
    return $\post(\vectheta|\vecs^o)$ as in \eqref{formula:agc-post}
    \caption{\texttt{Fine-grained learning}}
\end{algorithm}

The calculation of JSD is done numerically by a Riemann sum over $30^{K}$ equally spaced grid points with $K$ being the dimensionality of $\vectheta$. The region of the grids is defined by the minimal and maximal values of the samples from $P$ and $Q$ jointly. Before the JSD calculation, all distributions are further re-normalized so that they sum to unity over the grid.

To make the comparison sensible, each method has the same total number of simulations (budget) $N$ available. Note, however, that the different methods may use the simulation budget in different ways:
\begin{itemize}[leftmargin=*]
    \item For rejection ABC we pick the top 2000 samples closest to $\vecs^o$ for posterior estimation;
    \item For the two linear regression-based methods, regression ABC and GC-ABC, we pick the top $2000$ samples for posterior learning; 
    \item For NN-ABC and SNP-ABC we use all samples for learning, in line with their goal of modelling the relationship between $\vecs$ and $\vectheta$ without restriction to an $\epsilon$-ball. 
    \item For AGC-ABC, as mentioned before, we use the top 20\% of the samples in the coarse-grained phase while the top 2000 samples in the fine-grained phase. 
\end{itemize}
Both AGC-ABC and the SNP-ABC learn the posteriors in a coarse-to-fine fashion. For these two methods, we assign 20\%/80\% of the overall simulation budget $N$ to the coarse/fine phase respectively. Following the original literature \citep{mdn-abc1}, the neural networks in SNP-ABC have one/two hidden layers to model one/eight mixture of Gaussian in the two phases respectively, with 50 tanh units in each layer. For AGC-ABC, the neural network, which is considered in model selection, has two hidden layers with 128 and 16 sigmoid units each. This is also the same network used in NN-ABC. All neural networks in the experiments are trained by Adam \citep{ADAM} with its default setting.

\subsection{Results}
\begin{figure}[t!]
  \vspace{0.25cm}
  \centering
  \includegraphics[width=1.0\linewidth]{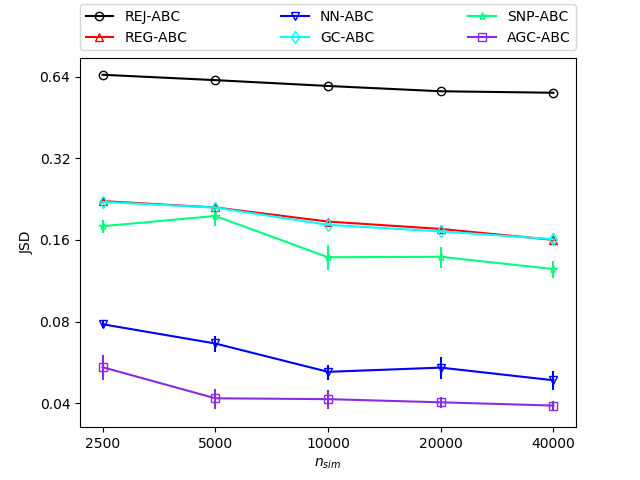}
  \caption{\label{figure:gc-jsd}The JSD average across 15 runs in the GC toy problem under different simulation budgets.}
\end{figure}
\paragraph{A toy problem} The first problem we study is a 2D Gaussian copula model:
\begin{align*}
  \hspace{-4ex}  \prob(\vecx|\vectheta) &= c_{\text{CG}}
        \Bigg(
         u_1, u_2;
         \begin{bmatrix}
          1, & \theta_3 \\
          \theta_3, & 1\\
         \end{bmatrix} \Bigg)
         \prod\limits_{k=1}^2 f_k(x_k;\theta_k) \\ 
    f_1(x_1;\theta_1) &= \text{Beta}(x_1; \theta_1, 2) \\
    f_2(x_2;\theta_2) &= \theta_2 \mathcal{N}(x_2; 1, 1) + (1-\theta_2) \mathcal{N}(x_2; 4, 0.25)
\label{formula:gc-model}
\end{align*}
where $u_k=F_k(x_k)$ is the value of the marginal CDF at $x_k$ and $c_{\text{CG}}(\cdot)$ is the aforementioned Gaussian copula density. The parameters of interest are $\vectheta = (\theta_1, \theta_2, \theta_3 )$ and the true parameters are $\vectheta^* = (6.0, 0.5, 0.6)$. We place a flat, independent prior on $\vectheta$: $\theta_1 \sim \mathcal{U}(0.5, 12.5)$, $\theta_2 \sim \mathcal{U}(0, 1)$, $\theta_3 \sim \mathcal{U}(0.4, 0.8)$. The summary statistics $\vecs$ are taken to be the union of (a) the 20 equally-spaced marginal quantiles and (b) the correlation between $z_1 = \Phi^{-1}(u_1)$, $ z_2 = \Phi^{-1}(u_2)$ of the generated data $\obs = \{\vecx^{(1)},\ldots, \vecx^{(200)} \}$. In this problem, the true posterior is available analytically.

Figure \ref{figure:gc-jsd} shows the JSD between the true and estimation posterior for the different methods (on log-scale, vertical lines indicate standard errors, each JSD is obtained by calculating the average of 15 runs for different observed data, the results shown in the figures below have the same setup). Expectedly given the model specification, AGC-ABC yields a lower JSD than the other methods for all simulation budgets. We further see that there is only a small difference in performance between AGC-ABC and NN-ABC, especially in large budget settings. A possible explanation is that the distribution of the residuals $p(\vecxi|\vecs)$ does only weakly depend on $\vecs$. To further investigate this, we compared the distribution of the residuals at $\vecs^o$ with the distribution $p_{\epsilon}(\vecxi)$ of the residuals within $\epsilon$-balls of increasing radius around $\vecs^o$. 
Table 1 shows the JSD values between the two distributions for different values of $\epsilon$.\footnote{For $\epsilon$ = 50\%, the JSD is 0.061.} It can be seen that the JSD between $p(\vecxi|\vecs^o)$ and $p_{\epsilon}(\vecxi)$ increases as $\epsilon$ increases but the changes are not large. This is in line with the small gain of AGC-ABC over NN-ABC. The supplementary material further compares the contour plots of $p(\vecxi|\vecs^o)$ and $p_{\epsilon}(\vecxi)$. 

\begin{table}[h]
    \centering
    \begin{tabular}{lcccc}
         \toprule
          $\epsilon$ (quantile) &  0.1\% & 1\% & 10\% & 25\% \\
         \hline
         $\text{JSD}(p(\vecxi|\vecs^o), p_{\epsilon}(\vecxi))$  \hspace{-0.33cm}  & 0.040 &  0.041 & 0.047 & 0.055  \\
         \hline
    \end{tabular}
    \caption{GC toy problem: The JSD values indicate a dependency of the residuals on $\vecs$.}
    \label{table:gc}
\end{table}

\paragraph{The M/G/1 queueing problem.} 
The M/G/1 queueing model is a real-world problem that describes the processing of incoming jobs in a single server system. Jobs in this model are served on a first-in-first-out (FIFO) basis. The whole data generating process in M/G/1 can be described as:
\[
\begin{split}
    s_i  &\sim \mathcal{U}(\theta_1, \theta_1 + \theta_2) \\
    v_i - v_{i-1} &\sim  \text{Exp}(\theta_3) \\
    d_i - d_{i-1} &= s_i + \text{max}(0, v_i - d_{i-1})
\end{split}
\]
where $s_i$, $v_i$, $d_i$ are the service time, visiting time and finish time of job $i$ respectively. The parameters of interest in this problem are $\vectheta = ( \theta_1, \theta_2, \theta_3 )$ and the true parameter values are $\vectheta^* = ( 1.0, 4.0, 0.2 )$. Again, we place a flat, independent prior on $\vectheta$: $\theta_1 \sim \mathcal{U}(0, 10)$, $\theta_2 \sim \mathcal{U}(2, 6)$, $\theta_3 \sim \mathcal{U}(0, 0.33)$. The observed data $\obs$ in this problem are given by the time intervals $x_i = d_i - d_{i-1}$ between the finish time of two consecutive jobs. 200 such intervals are observed so that $\obs = \{x_1, \ldots, x_{200} \}$. The summary statistics $\vecs^o$ are taken as the 16-equally spaced quantiles of $\obs$. These statistics are further preprocessed to have zero mean and unit variance on each dimension. 

\begin{figure}[t]
  \vspace{0.25cm}
  \centering
  \includegraphics[width=1.0\linewidth]{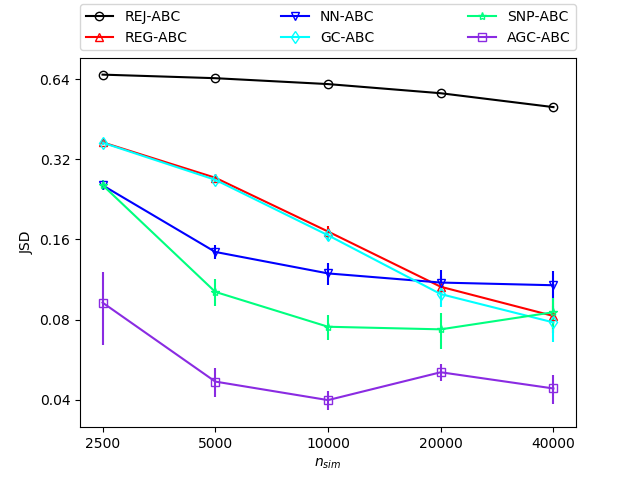}
  \caption{\label{figure:mg1-jsd} The JSD average across 15 runs in the M/G/1 problem under different simulation budgets.}
\end{figure}

Figure \ref{figure:mg1-jsd} shows the JSD of the different methods under various simulation budgets. We see that there is a larger gap between AGC-ABC and traditional regression ABC methods for small simulation budgets. This is because the underlying residual strongly depends on $\vecs$ in this problem; it is highly heterogeneous (see Table \ref{table:mg1} and the contour plots in the supplementary material). This also points to why NN-ABC does here not perform as good as AGC-ABC --- only modelling the dependency between the summary statistics and the scale (variance) of the residuals does not seem enough. SNP-ABC achieves very good performance compared to traditional ABC methods due to its heterogeneous modelling but it is still less accurate than AGC-ABC, which is natural since it requires more data to fit.

\begin{table}[h!]
    \centering
    \begin{tabular}{lcccc}
         \toprule
          $\epsilon$ (quantile) &  0.1\% & 1\% & 10\% & 25\% \\
         \hline
         $\text{JSD}(p(\vecxi|\vecs^o), p_{\epsilon}(\vecxi))$  \hspace{-0.33cm} &  0.047 & 0.112 & 0.159 & 0.194 \\
         \hline
    \end{tabular}
    \caption{M/G/1: JSD values between $p(\vecxi|\vecs^o)$ and $p_{\epsilon}(\vecxi)$ indicate a dependency of the residuals on $\vecs$.}
    \label{table:mg1}
\end{table}

\paragraph{MA(2) time series problem} The third problem we consider is the second order moving average model in time series analysis, known as MA(2). In this model, data are generated as follows
\begin{align*}
    x^{(t)} &= w^{(t)} + \theta_1 w^{(t-1)} + \theta_2 w^{(t-2)},
\end{align*}
where $x^{(t)}$ is the observation at time $t$ and $w^{(t)}$ is some unobservable standard normal noise. The parameters of interest are $\vectheta  = ( \theta_1, \theta_2 )$ and the true parameters are  $\vectheta^* = (0.6, 0.2)$. As in the previous problems, we place a flat and uniform prior on $\vectheta$: $\theta_1 \sim \mathcal{U}(0, 1)$, $\theta_2 \sim \mathcal{U}(0, 1)$. Following the ABC literature \citep{ma2-1, ma2-2,ma2-3}, we adopt the autocovariance with lag 1 and lag 2 as the summary statistics. These statistics are computed from a time series $\obs = (x^{(1)}, \ldots, x^{(200)})$ of length 200. The statistics are further preprocessed to have zero mean and unit variance for each dimension. 

\begin{figure}[t!]
  \vspace{0.25cm}
  \centering
  \includegraphics[width=1.0\linewidth]{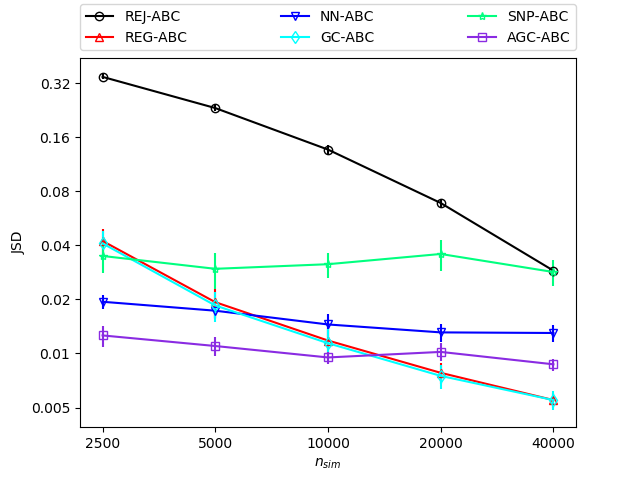}
   \caption{\label{figure:ma2-jsd} The JSD average across 15 runs in the MA(2) problem under different simulation budgets.}
\end{figure}

Figure \ref{figure:ma2-jsd} reports the JSD of the different methods under various simulation budgets. While AGC-ABC performs best for small budgets, the simpler REG-ABC method performs better in the larger budget case. This might be due to Gaussian copula being inadequate for the modelling of the residuals. However, the difference between AGC-ABC and REG-ABC is not large as evidenced by the contour plots in the supplementary materials. We also see that the 8-mixture of Gaussian construction in SNP-ABC might be not powerful enough for this problem, so that it incurs a bias which is not
eliminated as the simulation budget increases.

\begin{table}[h!]
    \centering
    \begin{tabular}{lcccc}
         \toprule
          $\epsilon$ (quantile) &  0.1\% & 1\% & 10\% & 25\% \\
         \hline
         $\text{JSD}(p(\vecxi|\vecs^o), p_{\epsilon}(\vecxi))$  \hspace{-0.33cm} &  0.004 & 0.004 & 0.006 & 0.010 \\
         \hline
    \end{tabular}
    \caption{MA(2): JSD values between $p(\vecxi|\vecs^o)$ and $p_{\epsilon}(\vecxi)$ indicate a dependency of the residuals on $\vecs$.}
    \label{table:ma2}
\end{table}

\paragraph{Lotka-Volterra problem} The last problem we consider is a stochastic dynamical system in biology that describes predator-prey dynamics. There are four possible events in the system: (a) a predator being born, (b) a predator dying, (c) a prey being born, (d) a prey being eaten by the predator. The events can be described probabilistically as
\[
    \begin{split}
          \Prob(X \to X+1 | \vectheta) &\propto \text{exp}(\theta_1) \cdot XY,   \\
          \Prob(X \to X-1 | \vectheta) &\propto \text{exp}(\theta_2) \cdot X, \\
          \Prob(Y \to Y+1 | \vectheta) &\propto \text{exp}(\theta_3) \cdot Y,  \\
          \Prob(Y \to Y-1 | \vectheta) &\propto \text{exp}(\theta_1) \cdot XY, 
    \end{split}
    \label{formula:lv-model}
\]
where $X, Y$ are the numbers of predators and preys respectively. For a more rigorous formulation see e.g.\ \citep{ss1}. The parameters of interest are $\vectheta = ( \theta_1, \theta_2, \theta_3 )$ and the true parameters are $\vectheta^* = ( \log(0.01), \log(0.5), \log(0.1) )$. The initial sizes of the populations are $( X^0, Y^0 ) =  (50, 100)$. While the likelihood is intractable, sampling from the model is possible \citep{g-algorithm}. We place a flat and independent prior on $\vectheta$:
$\theta_1 \sim \mathcal{U}(-5, -1)$, $\theta_2 \sim \mathcal{U}(-1, 1)$, $\theta_3 \sim \mathcal{U}(-1, 1)$. We simulate the system for a total of 8 time units and record the values of $X$ and $Y$ after every 0.2 time units. This yields two time series of length 41 being our observed data $\obs$. The summary statistics are  taken as the records between every two consecutive time points.

\begin{figure}[t!]
  \vspace{0.25cm}
  \centering
  \includegraphics[width=1.0\linewidth]{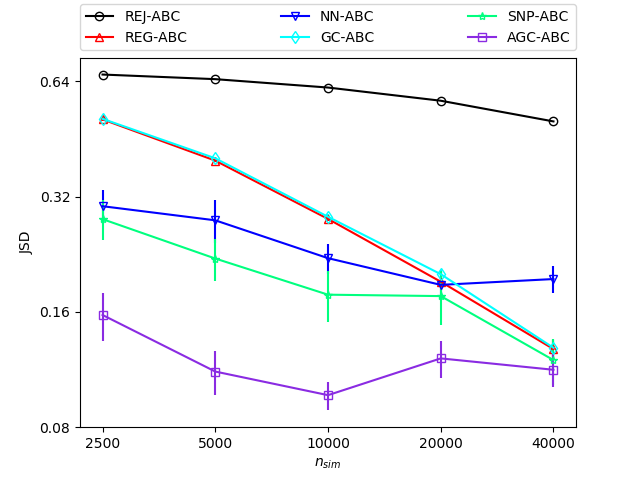}
    \caption{\label{figure:lv-jsd} The JSD average across 15 runs in the LV problem under different simulation budgets.}
\end{figure}

Figure \ref{figure:lv-jsd} shows the JSD of the different methods. The results show that AGC-ABC offers a better efficiency-accuracy trade-off than the other methods, with a clear performance gain for small simulation budgets. This is due to the fact that the underlying distribution $p(\vecxi|\vecs)$ is highly heterogeneous (see Table \ref{table:lv} and the the contour plots of $p(\vecxi|\vecs)$ in the supplementary material).

\begin{table}[h!]
    \centering
    \begin{tabular}{lcccc}
         \toprule
          $\epsilon$ (quantile) &  0.1\% & 1\% & 10\% & 25\% \\
         \hline
         $\text{JSD}(p(\vecxi|\vecs^o), p_{\epsilon}(\vecxi))$  \hspace{-0.33cm} &  0.089 & 0.108 & 0.149 & 0.194 \\
         \hline
    \end{tabular}
    \caption{LV: JSD values between $p(\vecxi|\vecs^o)$ and $p_{\epsilon}(\vecxi)$ indicate a dependency of the residuals on $\vecs$.}
    \label{table:lv}
\end{table}

\section{Conclusion}
We considered the problem of posterior density estimation when the likelihood is intractable but sampling from the model is possible. We proposed a new method for approximate Bayesian computation (ABC) that combines the basic ideas from two different lines of ABC research, namely regression ABC and sequential ABC. We found that the resulting algorithm strikes a good trade-off between accuracy and computational cost, being particularly effective in the regime of smaller simulation budgets. 

The motivation behind the proposed algorithm was the homogeneity assumption on the residuals that is required for regression ABC to work well. The proposed method takes a sequential approach by first generating training data so that the homogeneity assumption is better satisfied, and then models the data with the aid of a Gaussian copula and existing techniques from regression ABC. The method --- adaptive Gaussian copula ABC ---  can thus either be viewed as an adaptive version of classical regression ABC methods \citep{linRegABC, nonlinRegABC}, or a computationally cheaper version of recent sequential neural posterior approaches \citep{mdn-abc1, mdn-abc2}.

While Gaussian copula are powerful they are not silver bullets. Extending the Gaussian copula model to other statistical models such as mixture of Gaussian copula \citep{gc-abc2, mixturecopula} or vine copulas \citep{vinecopula, vinecopula2} may be a future research direction worth exploring.

\section*{Acknowledgement}
We would like to thank the anonymous reviewers for their insightful and fruitful suggestions. \\

\bibliography{main.bib}
\appendix
\onecolumn

\section{Contour plots of the residuals in each problem}

\subsection{Gaussian copula toy problem}

\begin{figure*}[h]
    \centering
    \subfigure[$p(\xi_1, \xi_2|\vecs^o)$]{
        \includegraphics[width=0.28\linewidth]{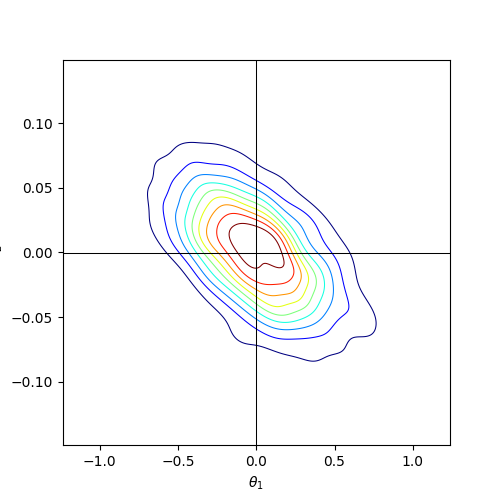}
    }
    \hspace{0.01\linewidth}
    \subfigure[$p(\xi_1, \xi_3|\vecs^o)$]{
        \includegraphics[width=0.28\linewidth]{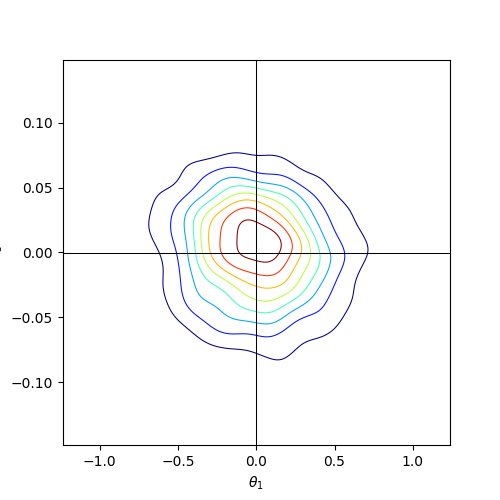}
    }
    \hspace{0.01\linewidth}
    \subfigure[$p(\xi_2, \xi_3|\vecs^o)$]{
        \includegraphics[width=0.28\linewidth]{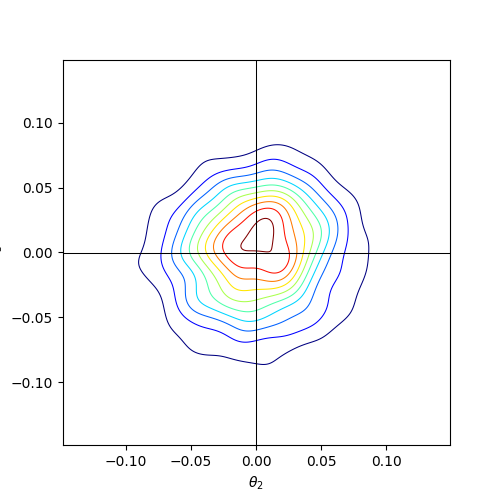}
    }
    \\
     \subfigure[$p_{\epsilon}(\xi_1, \xi_2)$, $\epsilon = 0.1\%$]{
        \includegraphics[width=0.28\linewidth]{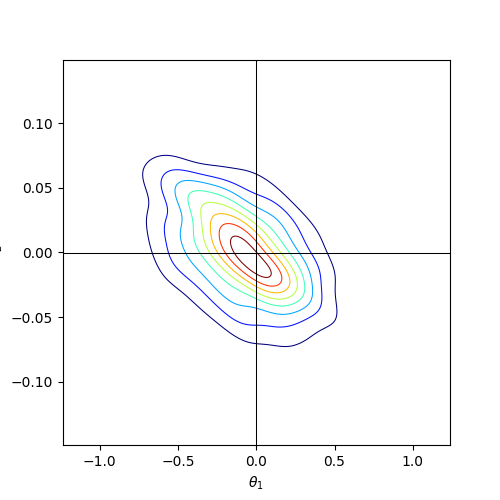}
    }
    \hspace{0.01\linewidth}
    \subfigure[$p_{\epsilon}(\xi_1, \xi_3)$, $\epsilon = 0.1\%$]{
        \includegraphics[width=0.28\linewidth]{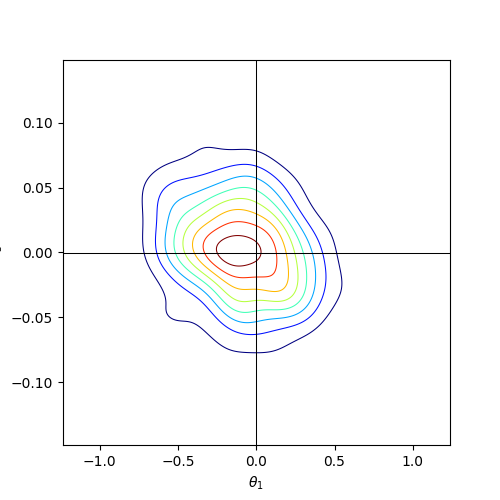}
    }
    \hspace{0.01\linewidth}
    \subfigure[$p_{\epsilon}(\xi_2, \xi_3)$, $\epsilon = 0.1\%$]{
        \includegraphics[width=0.28\linewidth]{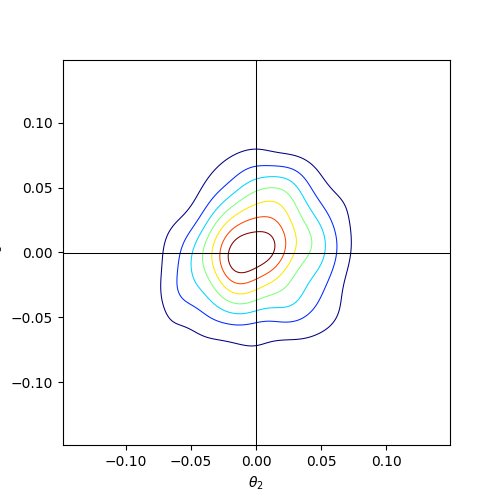}
    }
    \\
     \subfigure[$p_{\epsilon}(\xi_1, \xi_2)$, $\epsilon = 25\%$]{
        \includegraphics[width=0.28\linewidth]{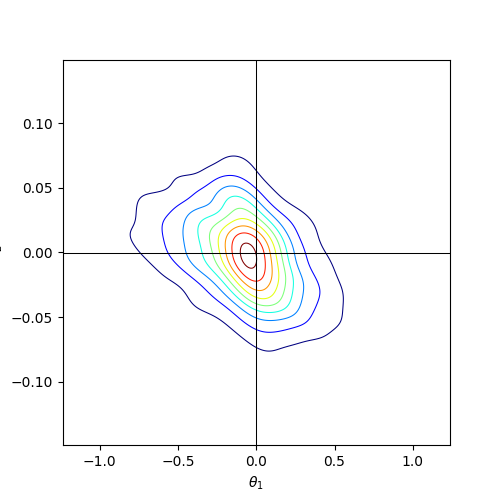}
    }
    \hspace{0.01\linewidth}
    \subfigure[$p_{\epsilon}(\xi_1, \xi_3)$, $\epsilon =  25\%$]{
        \includegraphics[width=0.28\linewidth]{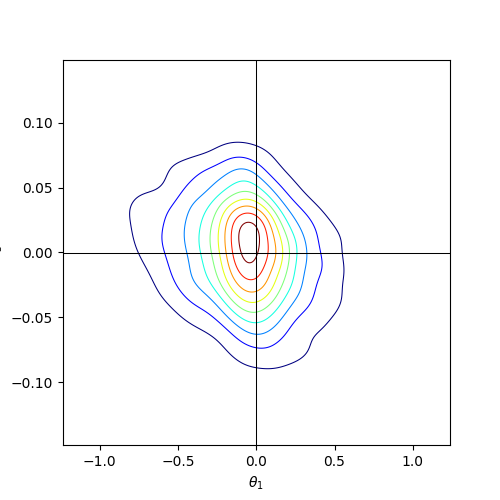}
    }
    \hspace{0.01\linewidth}
    \subfigure[$p_{\epsilon}(\xi_2, \xi_2)$, $\epsilon = 25\%$]{
        \includegraphics[width=0.28\linewidth]{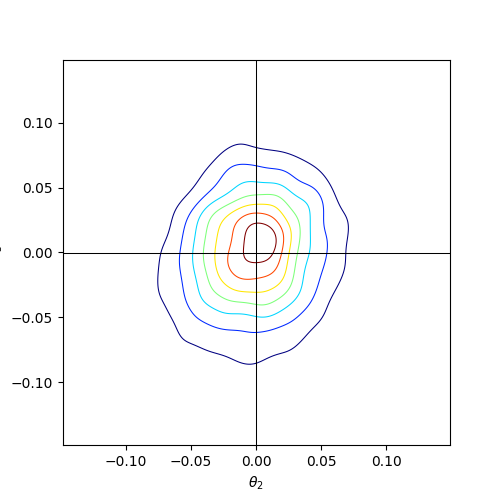}
    }
    \caption{Gaussian copula toy problem: visualizing the distribution of the residuals at $\vecs^o$ and the distribution $p_{\epsilon}(\vecxi)$ of the residuals within $\epsilon$-balls of different radii.}
    \label{figure:nn-comparison}
\end{figure*}

\clearpage

\subsection{M/G/1 problem}
\begin{figure*}[h]
    \centering
    \subfigure[$p(\xi_1, \xi_2|\vecs^o)$]{
        \includegraphics[width=0.28\linewidth]{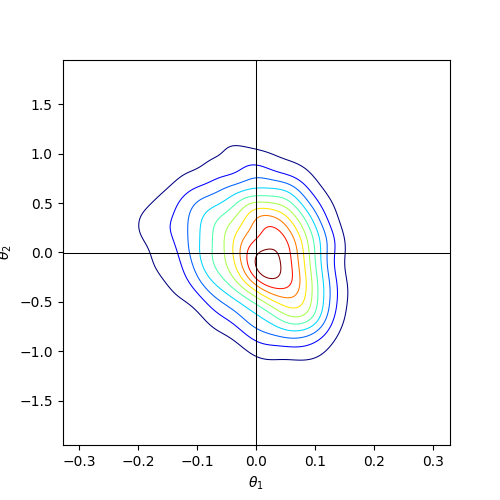}
    }
    \hspace{0.01\linewidth}
    \subfigure[$p(\xi_1, \xi_3|\vecs^o)$]{
        \includegraphics[width=0.28\linewidth]{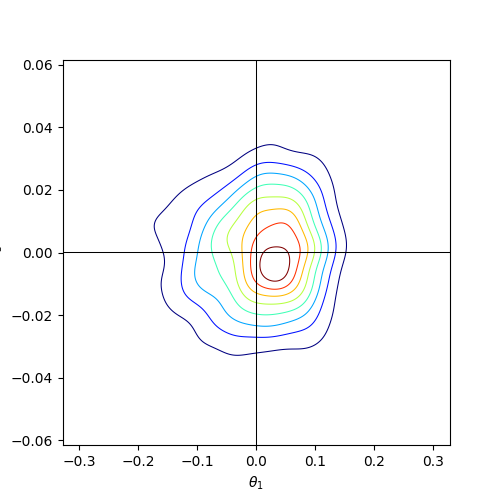}
    }
    \hspace{0.01\linewidth}
    \subfigure[$p(\xi_2, \xi_3|\vecs^o)$]{
        \includegraphics[width=0.28\linewidth]{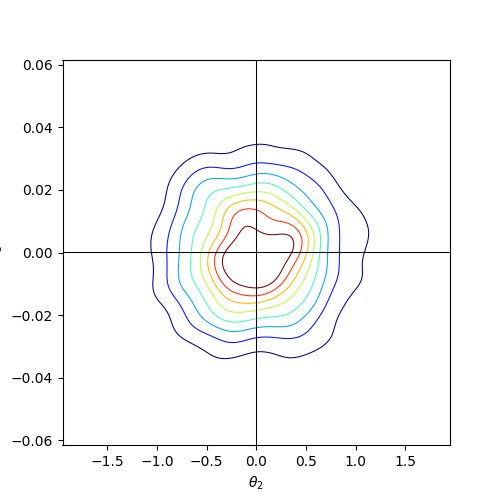}
    }
    \\
     \subfigure[$p_{\epsilon}(\xi_1, \xi_2)$, $\epsilon = 0.1\%$]{
        \includegraphics[width=0.28\linewidth]{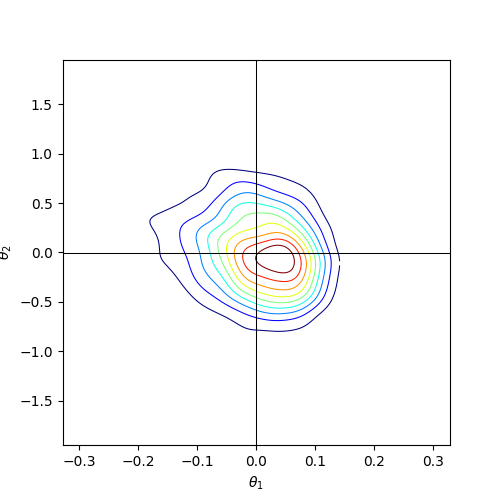}
    }
    \hspace{0.01\linewidth}
    \subfigure[$p_{\epsilon}(\xi_1, \xi_3)$, $\epsilon = 0.1\%$]{
        \includegraphics[width=0.28\linewidth]{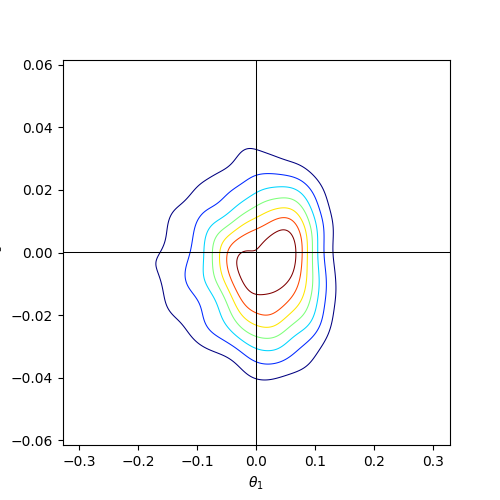}
    }
    \hspace{0.01\linewidth}
    \subfigure[$p_{\epsilon}(\xi_2, \xi_3)$, $\epsilon = 0.1\%$]{
        \includegraphics[width=0.28\linewidth]{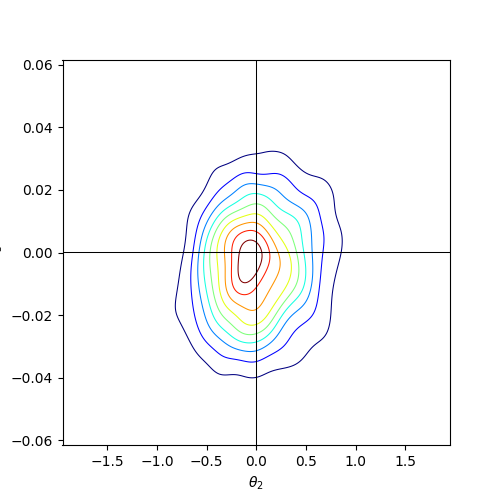}
    }
    \\
     \subfigure[$p_{\epsilon}(\xi_1, \xi_2)$, $\epsilon = 25\%$]{
        \includegraphics[width=0.28\linewidth]{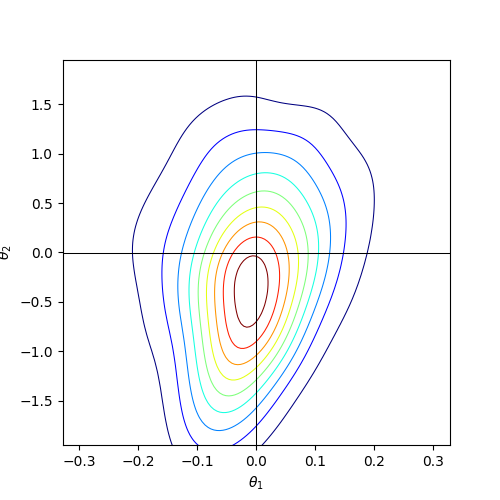}
    }
    \hspace{0.01\linewidth}
    \subfigure[$p_{\epsilon}(\xi_1, \xi_3)$, $\epsilon =  25\%$]{
        \includegraphics[width=0.28\linewidth]{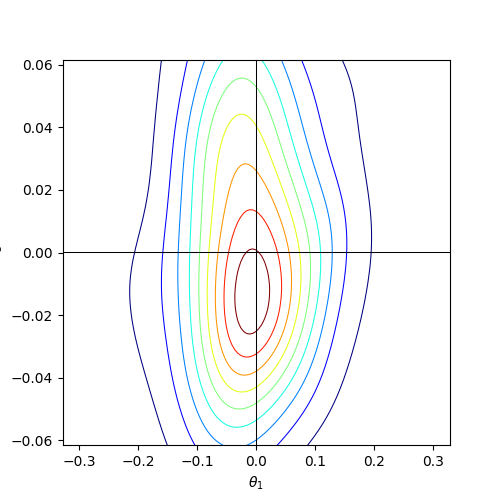}
    }
    \hspace{0.01\linewidth}
    \subfigure[$p_{\epsilon}(\xi_2, \xi_2)$, $\epsilon = 25\%$]{
        \includegraphics[width=0.28\linewidth]{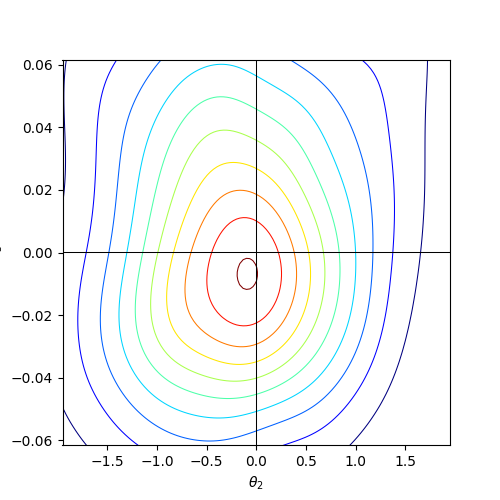}
    }
    \caption{M/G/1 problem: visualizing the distribution of the residuals at $\vecs^o$ and the distribution $p_{\epsilon}(\vecxi)$ of the residuals within $\epsilon$-balls of different radii.}
    \label{figure:nn-comparison}
\end{figure*}

\clearpage

\subsection{MA(2) problem}
\begin{figure*}[h]
    \centering
    \subfigure[$p(\xi_1, \xi_2|\vecs^o)$]{
        \includegraphics[width=0.28\linewidth]{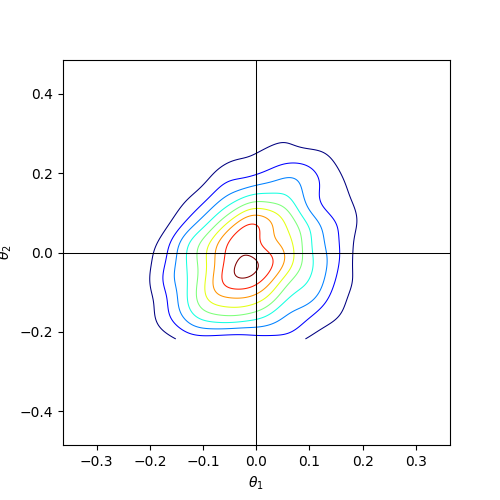}
    }
    \\
     \subfigure[$p_{\epsilon}(\xi_1, \xi_2)$, $\epsilon = 0.1\%$]{
        \includegraphics[width=0.28\linewidth]{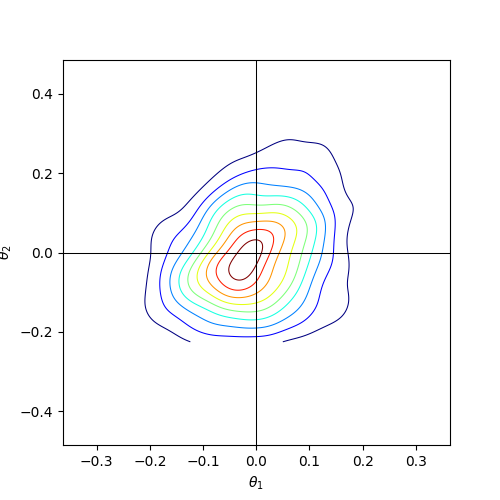}
    }
    \\
     \subfigure[$p_{\epsilon}(\xi_1, \xi_2)$, $\epsilon = 25\%$]{
        \includegraphics[width=0.28\linewidth]{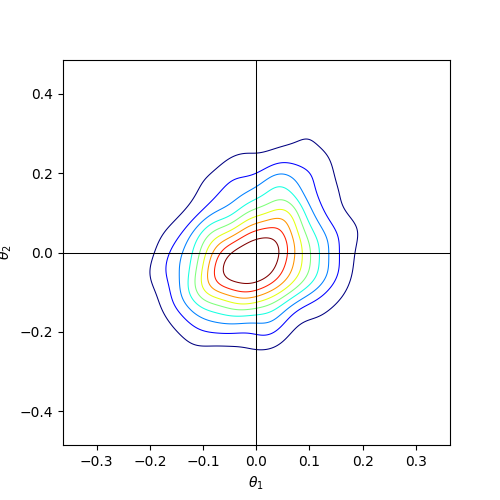}
    }
    \caption{MA(2): visualizing the distribution of the residuals at $\vecs^o$ and the distribution $p_{\epsilon}(\vecxi)$ of the residuals within $\epsilon$-balls of different radii.}
    \label{figure:nn-comparison}
\end{figure*}

\clearpage
\subsection{Lotka-Volterra problem}

\begin{figure*}[h]
    \centering
    \subfigure[$p(\xi_1, \xi_2|\vecs^o)$]{
        \includegraphics[width=0.28\linewidth]{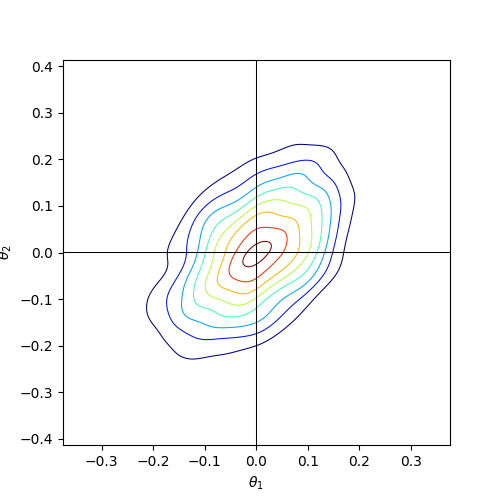}
    }
    \hspace{0.01\linewidth}
    \subfigure[$p(\xi_1, \xi_3|\vecs^o)$]{
        \includegraphics[width=0.28\linewidth]{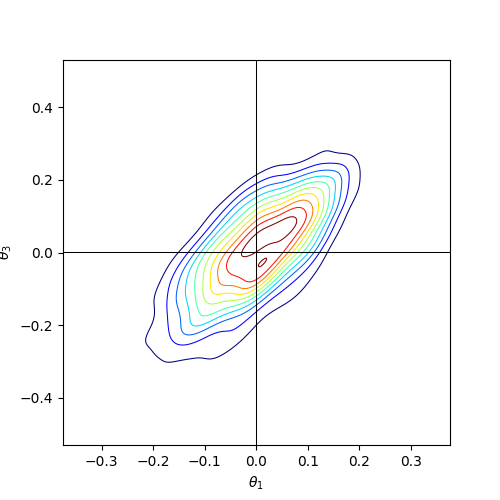}
    }
    \hspace{0.01\linewidth}
    \subfigure[$p(\xi_2, \xi_3|\vecs^o)$]{
        \includegraphics[width=0.28\linewidth]{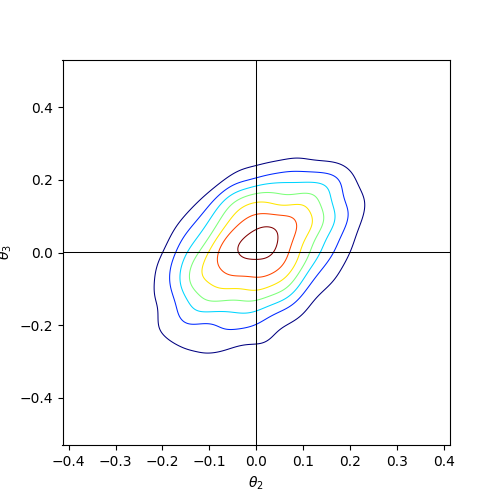}
    }
    \\
     \subfigure[$p_{\epsilon}(\xi_1, \xi_2)$, $\epsilon = 0.1\%$]{
        \includegraphics[width=0.28\linewidth]{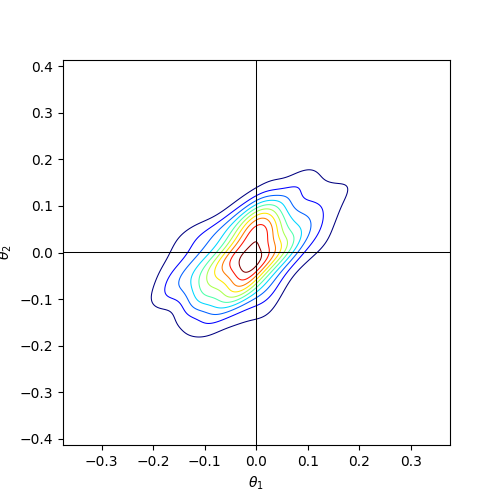}
    }
    \hspace{0.01\linewidth}
    \subfigure[$p_{\epsilon}(\xi_1, \xi_3)$, $\epsilon = 0.1\%$]{
        \includegraphics[width=0.28\linewidth]{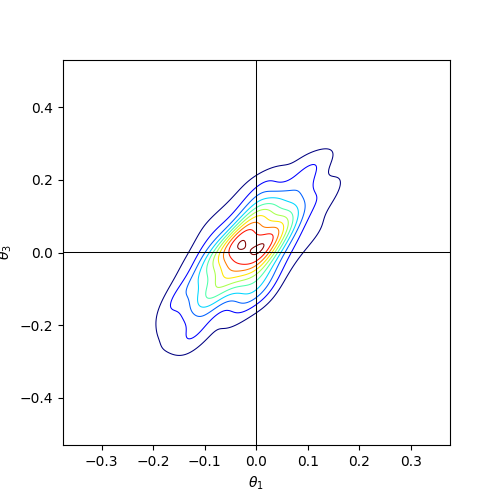}
    }
    \hspace{0.01\linewidth}
    \subfigure[$p_{\epsilon}(\xi_2, \xi_3)$, $\epsilon = 0.1\%$]{
        \includegraphics[width=0.28\linewidth]{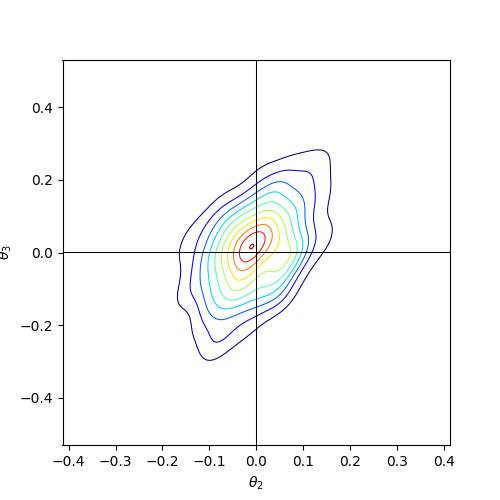}
    }
    \\
     \subfigure[$p_{\epsilon}(\xi_1, \xi_2)$, $\epsilon = 25\%$]{
        \includegraphics[width=0.28\linewidth]{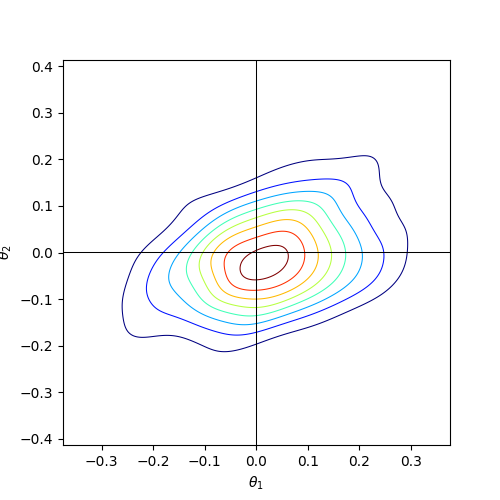}
    }
    \hspace{0.01\linewidth}
    \subfigure[$p_{\epsilon}(\xi_1, \xi_3)$, $\epsilon =  25\%$]{
        \includegraphics[width=0.28\linewidth]{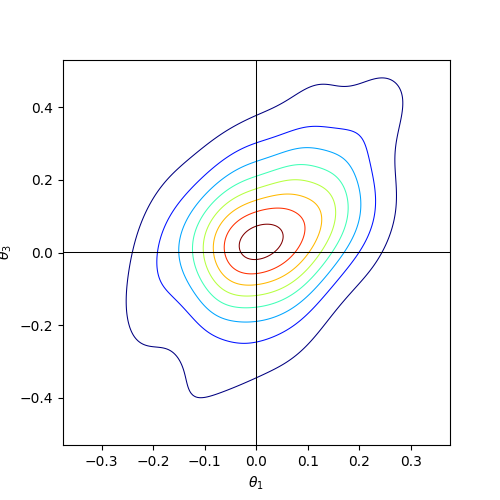}
    }
    \hspace{0.01\linewidth}
    \subfigure[$p_{\epsilon}(\xi_2, \xi_2)$, $\epsilon = 25\%$]{
        \includegraphics[width=0.28\linewidth]{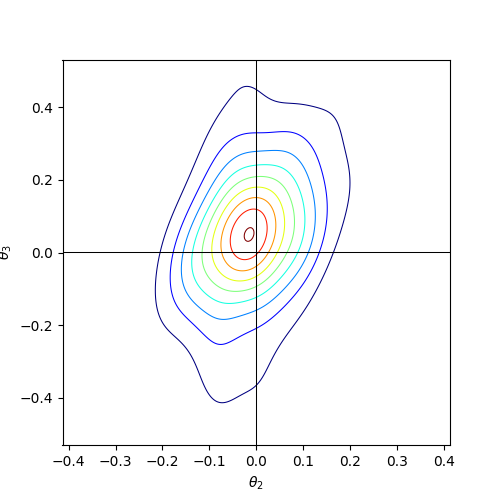}
    }
    \caption{Lotka-Volterra problem: visualizing the distribution of the residuals at $\vecs^o$ and the distribution $p_{\epsilon}(\vecxi)$ of the residuals within $\epsilon$-balls of different radii.}
    \label{figure:nn-comparison}
\end{figure*}
\clearpage

\section{Contour plots of approximate posterior learned by each method in MA(2)}

\begin{figure*}[h]
    \centering
    \subfigure[True posterior]{
        \includegraphics[width=0.28\linewidth]{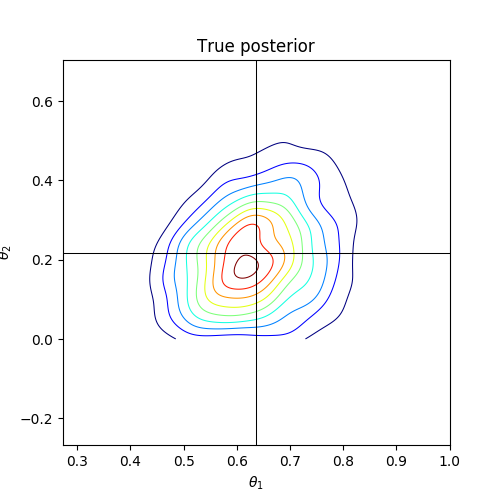}
    }
    \\
     \subfigure[REG-ABC posterior]{
        \includegraphics[width=0.28\linewidth]{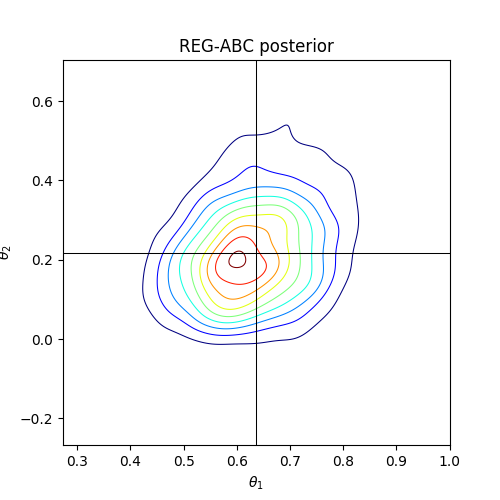}
    }
    \\
     \subfigure[AGC-ABC posterior]{
        \includegraphics[width=0.28\linewidth]{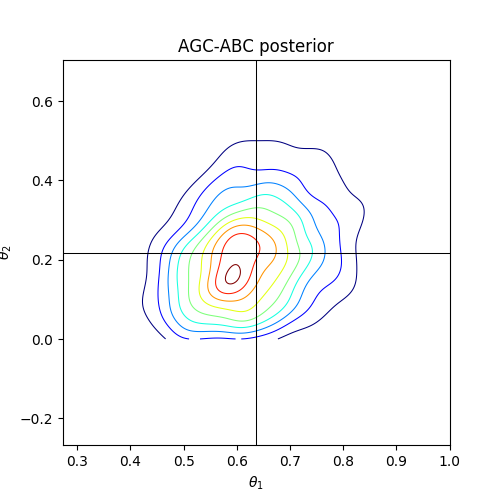}
    }
    \caption{MA2 problem: the contour plots of the posteriors learned in each method with simulation budget  $N=2,500$.}
\end{figure*}

\end{document}